\documentclass[10pt,letterpaper]{article}
\usepackage[top=0.85in,left=2.75in,footskip=0.75in]{geometry}

\usepackage{changepage}

\usepackage[utf8]{inputenc}

\usepackage{textcomp,marvosym}

\usepackage{fixltx2e}

\usepackage{amsmath,amssymb}

\usepackage{cite}

\usepackage{nameref,graphicx,epstopdf}

\usepackage{microtype}
\DisableLigatures[f]{encoding = *, family = * }

\usepackage{rotating}

\raggedright
\usepackage[aboveskip=1pt,labelfont=bf,labelsep=period,justification=raggedright,singlelinecheck=off]{caption}

\makeatletter
\renewcommand{\@biblabel}[1]{\quad#1.}
\makeatother

\date{}

\usepackage{lastpage,fancyhdr,graphicx}
\usepackage{epstopdf}
\fancyheadoffset[L]{2.25in}
\fancyfootoffset[L]{2.25in}
\begin{document}
\vspace*{0.35in}

\begin{flushleft}
{\Large
\textbf\newline{Social and Spatial Clustering of People at Humanity's Largest Gathering}
}
\newline
\\
Ian Barnett\textsuperscript{1},
Tarun Khanna\textsuperscript{2},
Jukka-Pekka Onnela\textsuperscript{1,*}
\\
\bigskip
\bf{1} Department of Biostatistics, Harvard University, Boston, MA, USA
\\
\bf{2} Harvard Business School, Boston, MA, USA
\\
\bigskip

%
%





* onnela@hsph.harvard.edu

\end{flushleft}
\section*{Abstract}
Macroscopic behavior of scientific and societal systems results from the aggregation of microscopic behaviors of their constituent elements, but connecting the macroscopic with the microscopic in human behavior has traditionally been difficult. Manifestations of homophily, the notion that individuals tend to interact with others who resemble them, have been observed in many small and intermediate size settings. However, whether this behavior translates to truly macroscopic levels, and what its consequences may be, remains unknown. Here, we use call detail records (CDRs) to examine the population dynamics and manifestations of social and spatial homophily at a macroscopic level among the residents of 23 states of India at the Kumbh Mela, a 3-month-long Hindu festival. We estimate that the festival was attended by 61 million people, making it the largest gathering in the history of humanity. While we find strong overall evidence for both types of homophily for residents of different states, participants from low-representation states show considerably stronger propensity for both social and spatial homophily than those from high-representation states. These manifestations of homophily are amplified on crowded days, such as the peak day of the festival, which we estimate was attended by 25 million people. Our findings confirm that homophily, which here likely arises from social influence, permeates all scales of human behavior.


\section*{Introduction}
When the behavior of each individual in a group is dependent on their interactions with others around them, the collective behavior of the group as a whole can be surprisingly different from what would be expected by simply extrapolating off that of the individual \cite{schelling2006micromotives,strandburg2015shared,helbing2000simulating}. In particular, people think and behave differently in crowds than in small scale settings\cite{le1897crowd,park1972crowd,blumer1946elementary}, and this crowd behavior can occasionally lead to tragic events and even human stampedes\cite{ngai2009human,greenough2013kumbh,maclean2003power}. Individuals tend to form groups spontaneously and engage in collective decision-making outside of such dramatic events as well, but the nature of this type of herding--and the extent to which it happens--depends on how outnumbered the group is compared to the reference population. For example, friendship networks of adolescents demonstrate greater social homophily if they are in the minority \cite{gonzalez2007community}, whereas majority members do not share this preference\cite{vermeij2009ethnic}. This phenomenon is in line with the description by Simmel who argued that individuals ``resist being leveled'' in a crowd\cite{simmel1903metropolis}. If, however, the minority group is too small to form an independent community, it is possible for the minority to show heterophily rather than homophily\cite{currarini2009economic}. This finding highlights the importance of the surrounding social context, in particular the relative size of the group. Social homophily can also lead to spatial homophily and thereby give rise to segregation\cite{schelling1971dynamic, hatna2014combining}. 

While the term homophily is used to mean different things, we use it here to refer to the tendency for people who are similar to be associated with one another regardless of the mechanism that causes this association. This use of the term is distinct from quantifying homophily by the frequency of associations among similar people, since people in the majority will have a greater frequency of associations with others in the majority simply due to having more opportunities for forming them\cite{mcpherson2001birds,hallinan1985effects}. While several studies have investigated homophily of racial groups on smaller scales, we explore how such homophilous tendencies might persist on a much larger macroscopic scale. The behavior of individuals in a classroom cannot be used to extrapolate onto the behavior of those packed into a crowd of millions.

The Kumbh Mela is a religious Hindu festival that has been celebrated for hundreds of years\cite{mehrotra2015kumbh}, and the 2013 Kumbh Mela, organized in Allahabad, stands out from all others today and throughout history due to its magnitude. As it is infeasible to collect demographic data from millions of participants, we turned to call detail records (CDRs) that have been used to investigate social networks, mobility patterns, and other massive events  \cite{blondel2015survey,onnela2007structure,onnela2007analysis,gonzalez2008understanding,wesolowski2012quantifying,aleissawired}. Cell phone operators routinely maintain records of communication events, mainly phone calls and text messages, for billing and research purposes. These communication metadata, at minimum, keep track of who contacts whom, when, and for how long (voice calls only). Using these call detail records (CDRs), we first estimate the attendance of each of 23 states of India at the event before investigating the relationship between a state's attendance and the degree of both social homophily and spatial homophily amongst its attendees.


\section*{Methods}

\subsection*{Data description}

We had access to CDRs\footnote{Only summary statistics from the CDRs were provided to us: social network information and daily customer counts at various cell towers located at the Kumbh. Caller IDs were anonymized, and no individual-level characteristics were provided to us aside from billing area codes and whether or not a prepaid or postpaid plan was used.} for one Indian operator for the period from January 1 to March 31, 2013. This dataset contains records of 146 million (145,736,764) texts and 245 million (245,252,102) calls for a total of 390 million (390,988,866) communication events. Given the logistical impossibility of collecting demographic, linguistic, or cultural attributes of Kumbh participants at scale, we based our investigation of homophily on a marker that acts as a proxy for these covariates, namely, cell phone area codes. The area codes correspond to different states\footnote{Though officially India has more than 23 states, we adhere instead to the 23 functional state divisions used by the service provider.} of India, and as a result of India's States Reorganization Act of 1956 these divisions summarize demographic variability along linguistic origin, ethnic agglomeration, and preexisting social bonds and boundaries. While CDRs readily lend themselves to studying social networks and social homophily, to investigate spatial homophily we additionally acquired access to the cell tower IDs at the Kumbh venue.

Combined with the latitude and longitude of each of the 207 towers at the site\footnote{In anticipation of the large influx of people at the Kumbh, temporary infrastructure was brought into the venue prior to the start of the festival so as to provide sufficient coverage for the large number of expected cell phone users.}, we were able to infer the caller's location (at the time of phone-based communication) with relatively high spatial resolution. The grid that divides the Kumbh site into regions around each cell tower, called the Voronoi tessellation, groups all points on the map closest to each cell tower. The birds-eye view of Allahabad in Fig. 1 shows the estimated attendance on one of the busiest and most favorable days for ritual bathing in the Ganges river.

\vspace{1pc}\noindent\textbf{Figure 1. Cell phone usage around the cell towers at the Kumbh during its busiest day.} The heat map polygons represent the Voronoi tessellation around the cell towers that occupied the site of the Kumbh Mela event in Allahabad, India. Cell towers with no activity are removed from the analysis and their Voronoi cells are merged into neighboring active cell towers. Map data used to produce the river traces: Google, DigitalGlobe.
\vspace{1pc}

\subsection*{Attendance Estimation}

Extrapolating population measures from CDRs has become feasible in recent years due to the rapid increase in the prevalence of cell phones. While CDRs provide raw counts of cell phone users, to estimate attendance, these numbers need to be adjusted by (i) overall prevalence of cell phones in India, (ii) the state-specific market shares of our provider, (iii) the probability of daily use for a person known to be present at the venue, and (iv) the probability of phone non-use during a person's entire stay at the venue. First, regarding overall phone prevalence, $71.3\%$ of people in India had a wireless subscription in 2013 \cite{TRAIreport}.

Second, regarding market share, the number of unique handsets are counted on a daily basis for each of 23 distinct states of India (Table S1), as defined by the service provider, and each count is extrapolated from the service provider's market share in the given states. The service provider's market share varies widely state by state (range $13.7\%$, $42.6\%$). It is important to use state-specific market share, because if average market share is used instead, the state-specific attendance counts can be off by more than a factor of $2$. These handset counts are added together for each day before extrapolating to the general population.

Third, regarding daily use, it is likely that many Kumbh attendees who use their phone at least once do not use their phone every day while at the festival. If not addressed, this would bias our population estimate downwards. By tracking phone activity, length of stay can be estimated based on the time period a person's phone is active while at the Kumbh. Based on this, we estimate the percentage of customers who use their phone on any given day during their stay conditional on them using their phone at least once during their stay to be $40.4\%$. (Note that this quantity applies to daily estimates, not to cumulative estimates. See \nameref{S1_text}.)

Fourth, regarding non-use, the probability of a person not using his or her phone during the entire stay at the venue is difficult to account for; these individuals are not visible in the observed data, and yet the proportion of non-users could potentially be substantial given that many visitors from outside regions would have to pay roaming fees, which likely leads them to minimize their phone use. To overcome this difficulty, we first examine four available daily population projections\cite{projections}, each for a different day, and calibrate the proportion of non-users such that our resulting daily estimate for that same day is most consistent with the four daily projections. We obtain an estimate of $40.6\%$ for non-use (coincidentally similar to $40.4\%$ obtained above for daily use) and we use this estimate to adjust both cumulative and daily attendance.

\subsection*{Social Homophily}

A social network is constructed between customers who used their phone at the Kumbh. A network edge is assumed between any two people who communicated with one another at any point over the course of the Kumbh. To study how a state's extent of social homophily is related to its level of representation, defined as the number of people present from the state divided by the total Kumbh attendance, we select a measure that results in consistent estimates of homophily regardless of state representation. The measure of social homophily considered in refs.~\cite{coleman1958relational,currarini2009economic} applied to our setting would define homophily for any given state as the proportion of ties that involve two participants from that state, but due to measuring absolute differences instead of relative differences, the homophily for states with small representation would be biased downwards due to their small proportions. A standard stochastic block model (SBM) approach\cite{holland1983stochastic} applied to our setting would assume an equal likelihood of forming network ties between any two participants from the same state. However, if this model is misspecified and there exist additional social structure within each state (within each block), as is almost certainly the case, then this approach is likely biased in the opposite direction and overestimates the social homophily in states with lower representation.\footnote{To see the reason for this, consider the case where state A sends only a single group of friends to the Kumbh, whereas state B sends 100 different groups of friends. A random pair selected from state A will have a much higher likelihood of being friends than will a random pair from state B, even if social homophily is equally strong within the friendship groups of the two states.} The biases of both these methods are discussed in further detail in \nameref{S1_text}. 

To circumvent these problems, we shift our focus from dyads to same-state connected triples, sets of three nodes from the same state that are connected either by two edges, resulting in an open triple, or three edges, resulting in a closed triple. The rationale behind this choice is that the three nodes in a connected triple can be assumed to belong to the same social group whether the triple is open or closed. By considering the propensity for same-state connected triples to be closed, we can gain insight into how densely connected the social groups are in which these triples are embedded. This approach is a way of sampling pairs of nodes from the same social group even when the social groups themselves are unobserved. The proportion of triples that are closed provides a natural measure of social homophily (see Fig. 2). This measure is commonly referred to as the global clustering coefficient or the transitivity index \cite{wasserman1994social} calculated over each state-specific network. 
Ignoring residents from the local state whose phone use is likely different from all other states\footnote{When studying social homophily we ignore the attendees from the local state where the Kumbh is held, eastern Uttar Pradesh, because the social behavior of the locals is likely not comparable to those from the other 22 states. While visitors from other states are all present for the same purpose of participation in the Kumbh, this is not true for the locals, many of whom were employed to help run the Kumbh in various roles. Outsider phone usage will likely be exclusively for coordinating purposes at the event, due to the cost of roaming calls. On the other hand, locals use their phones much more freely and for everyday purposes.}, there are 1,630,553 connected triples in the full Kumbh social network. 

\vspace{1pc}\noindent \textbf{Figure 2. Schematics of homophily measures (A) and call detail records (B).} For homophily measures (\textbf{A}), the three dotted lines represent spatial boundaries for the Voronoi tessellation around the cell towers, separating the shaded region into three Voronoi cells, in two (a low and high homophily) examples. The solid lines denote which nodes are in communication in the social network, either through voice call or text message. In the context of spatial homophily, two nodes are considered nearby if and only if they both are in the same spatial region (Voronoi cell) on the same day. The size of Voronoi cells range from as small as a $1/4 \mbox{km}^2$ to as large as $20 \mbox{km}^2$. For the call detail records (\textbf{B}), analysis of spatial homophily uses all pairwise communication events involving at least one customer of our operator who is present at the Kumbh, whereas analysis of social homophily only considers the ties between customers of our operator.
\vspace{1pc}

Letting $C_{ijk}=1$ if the $(i,j,k)$ triple is closed and $C_{ijk}=0$ if it is open, and let $R_{ijk}$ be the state of the three nodes in the triple, with $W_{r}$ as the proportion of the total cumulative Kumbh population by March 31, 2013, that belongs to state $r=1,\dots,23$. Across the 22 non-local states, $W_{r}$ ranges from $0.018\%$ to $7.45\%$, thus varying over  2.5 orders of magnitude. We fit the following regression model over all connected triples:
\begin{equation}\label{logisticmodel}
\mbox{logit}(\mbox{pr}(C_{ijk}=1)) = \beta_0 + \beta_1 \log_{10}W_{R_{ijk}}
\end{equation}

The model requires independence between observations for accurate inference, and because the same individual can be involved in multiple triples, this independence does not hold. The estimate $\hat{\beta}_1$ from \eqref{logisticmodel} is still unbiased, but its standard error and the $P$-value for the two-sided test of the null hypothesis $\beta_1=0$ will not be correct if this dependence is ignored. Taking advantage of the large sample size, for accurate inference we select a random subset of triples where we do not allow the same individual to appear in more than one triple. 

\subsection*{Spatial Homophily} 
   
Let $n_{crd}$ be the number of customers near cell tower $c$ from state $r$ on day $d$ of the Kumbh, and let $N_{rd} = \sum_{c=1}^Cn_{crd}$ be the total number of customers from state $r$ at the Kumbh on day $d$, where the sum is taken over all $C$ cell towers. To avoid double-counting, if a person uses multiple cell towers on the same day, only the first cell tower is recorded. The probability that any two given individuals from the state $r$ are nearby on the day $d$ is:
\begin{equation}\label{nearbyprob}
p_{rd}=\frac{1}{N_{rd}}\sum_{c=1}^C n_{crd}\frac{n_{crd}-1}{N_{rd}-1}
\end{equation}
Here two people are defined to be ``nearby'' on a particular day when they are both located in the same Voronoi cell on that day, using the cell tower designation mentioned above. The intuition behind equation \ref{nearbyprob} is that, given the location of one person, the probability a different randomly selected person from their state is in the same Voronoi cell is $(n_{crd}-1)/(N_{rd}-1)$. The probability in equation \ref{nearbyprob} has the desirable property of not scaling with state representation $W_r$ if spatial homophily is kept constant.\footnote{To see this, suppose that we hold constant how a particular state's attendees are spread out over the cell towers of the Kumbh, i.e. suppose we fix $n_{crd}/N_{rd}$. If we then increase the number of people present at the Kumbh from that state, $p_{rd}$ will stay essentially unchanged with a negligible increase, because $(a\cdot n_{crd}-1)/(a\cdot N_{rd}-1) > (n_{crd}-1)/(N_{rd}-1)$ for any $a>1$.} This property is essential if we wish to evaluate the relationship between spatial homophily and state attendance/representation. Finally, let $Q_{r}^{A} = \sum_{d=1}^{90}p_{rd}/90$ be the probability that any two given individuals from state $r$ are nearby averaged over all 90 days.

To evaluate busy, or high volume, days, we consider the three days with the highest attendance. We grouped each of these three days together along with the two days that preceded each and the two days that followed each, leading to a set of 15 days we labeled as high volume days. The remaining 75 days were grouped together to form the set of low volume days. We let $Q_r^{H}$ be the average of the $p_{rd}$ over the high volume days,  $Q_r^{L}$ be the average of the $p_{rd}$ over the low volume days, and we defined $Q_r^{D} = Q_r^{H}/Q_r^{L}$ to be the ratio of spatial homophily when comparing high volume days to low volume days.

\section*{Results}

\subsection*{Attendance Estimation}

Since the extent of homophily for any given group can depend on the relative size of that group compared to others, we first estimate daily and cumulative attendance for participants from each state which can then simply be added up to obtain overall attendance estimates. Existing estimates of the Kumbh's attendance vary widely and most are obtained with heavy extrapolation based on rough head counts combined with the rate of flow at high traffic points leading to the Kumbh venue \cite{wsjarticle}. These estimates have the limitation that they only look at the primary entrances into the Kumbh and ignore traffic flow from secondary entrances. And while daily estimates can be inferred from traffic flow or satellite images, cumulative attendance is more difficult to obtain, because a satellite image cannot tell if the same people are present for many weeks, or if people stay only a short time before leaving to be replaced by newcomers. 

Our estimates for the total daily and cumulative attendance are shown in Fig. 3. They clearly show a spike of attendance on each of the Kumbh's three primary bathing days. These days hold special religious significance and bathing on these days is seen to be particularly auspicious.  Based on the above numbers, we estimate the peak daily attendance of the 2013 Kumbh on February 10th to be $25$ million, and the total cumulative attendance from January 1 to March 31 to be $60.6$ million, which suggests that the event was the largest recorded gathering in humanity's history. A sensitivity analysis in Fig. 3 shows the cumulative attendance if the percent of customers that are non-users is varied from the estimated $40.6\%$. For example, if the percent of customers that are non-users is $45\%$, then the cumulative attendance sinks to $54$ million, whereas if the percent of customers that are non-users is $35\%$, then the cumulative attendance rises to $69$ million.

\vspace{1pc}\noindent\textbf{Figure 3. Estimates for daily and cumulative attendance at the Kumbh.} The cumulative (\textbf{A}) and daily (\textbf{B}) attendance at the Kumbh is estimated from January 1st, 2013, to March 30th, 2013. Daily estimates  are the number of unique handsets used extrapolated by the (i) the national prevalence of cell phones, (ii) state-specific market share of the service provider, (iii) the likelihood of inactivity on a daily basis, and (iv) the proportion of individuals who never use their phone (non-users). Cumulative estimates are extrapolated only by (i), (ii), and (iv), which accounts for the apparent difference between daily and cumulative counts on January 1st. The sensitivity of total cumulative attendance to changes in (iv) shows the importance of accounting for this form of censoring in the data \textbf{(C)}. The curve plotted is $f(x)=c/x$, where $c=24467257$.
\vspace{1pc}

\subsection*{Social Homophily}

We investigate social homophily among the residents of the 23 states, using state-specific attendance estimates, by constructing a social network of Kumbh attendees. The network nodes correspond to people and edges correspond to one or more pairwise communication events between people. Note that only communication events involving the service provider's customers present at the Kumbh venue are observed (see Fig. 2), and both parties must be customers of the provider to be included in the network so that their state of residence can be ascertained. The resulting network contains 2,130,463 nodes and 8,204,602 ties. The network is constructed using the full three month period using both text and call information combined because otherwise the network would become too sparse if segmented.

When there is strong social homophily in a state, the connected triples in the social network among attendees from that state will have an increased likelihood of being closed. After fitting model \eqref{logisticmodel} we find that there is strong negative association between social homophily and state representation. The model fit has an estimate of $\hat{\beta}_1=-0.208$, $95\%$ CI $(-0.259,-0.157)$, implying that a ten-fold increase in $W_r$ corresponds to an $81\%$ decrease in the expected proportion of closed triples. The analysis restricted to a subset of independent triples yields a $P$-value less than $10^{-20}$ and this significance remains robust to the subset selected. This analysis reduces sample size and sacrifices some statistical power by looking only at a subset of independent triples in order to allow for accurate statistical inference. Even then, the $P$-value remains very significant, providing strong evidence that minority states at the Kumbh tend to show significantly greater social homophily as compared to well represented states.

\subsection*{Spatial Homophily}

Does the finding of heavily outnumbered states being more tightly-knit in their social networks apply to spatial homophily as well? We use our knowledge of which cell tower is used by a caller to approximate caller location. Let $Q_r^{A}$ be the probability that any two given individuals from state $r$ are physically nearby averaged over all 90 days of the Kumbh. The $Q_r^{A}$ and their confidence intervals are illustrated in Fig. 4, with $Q_{r}^{A}$ ranging between $0.0025$ and $0.018$, reflecting over a 7-fold difference in the propensity for spatial homophily across states, with a mean value of $0.013$. States with low representation tend to be more spatially homophilous than states with high representation. In contrast, the local people from the eastern Uttar Pradesh, where the Kumbh Mela takes place, alone make up a majority at the Kumbh, and they show significantly less spatial homophily. Overall, there is a strong negative correlation (Pearson's $\rho=-0.54$) between spatial homophily ($Q_j^{A}$) and average logarithmic daily representation at the Kumbh.

\vspace{1pc}\noindent\textbf{Figure 4. The spatial homophily and representation of the 23 mainland states of India at the Kumbh.} The point estimates and $95\%$ confidence intervals for $Q_r^{A}$, the probability that any two given customers from state $r$ are physically close to one another, (\textbf{A}) and $Q_r^{D}$, the relative increase of state $r$'s spatial homophily on busy days compared to normal days, (\textbf{B}), both demonstrate an inverse relationship with state representation. The states have been ranked first by representation at the Kumbh (\textbf{C}) and then by degree of spatial homophily (\textbf{D}) (see \nameref{S1_text} for the list of state names). The heat map colors correspond to the rankings. The yellow star is the city of Allahabad, the location of the 2013 Kumbh Mela. The near inversion of colors when comparing the two panels demonstrates a clear negative association between state representation and spatial homophily.

\vspace{1pc}

The average spatial homophily $Q_r^{A}$ above was computed over the full three-month period, but it is conceivable that spatial homophily is a dynamic characteristic that varies from day to day, reflecting the changing compositions of different social groups. We conjectured that the extent of spatial homophily might be different on the three primary bathing days of  February 10, February 15, and March 10 as compared to the other less crowded days. To test this, we define $Q_r^{D}$ to be the ratio of spatial homophily on crowded, high volume, days relative to spatial homophily on lower attendance days for state $r$.

Fig. 4 shows that states with low representation tend to have a greater increase in spatial homophily on the high volume days. Participants from these underrepresented states appear particularly sensitive to increase in crowds, and they seem to group together more closely as the crowds build up. Some of the states with high representation are more robust to changes in the crowd size. In fact, there were seven states that had the opposite effect (though these effects were quite mild in comparison). There is a gap between the top four most represented states at the Kumbh (Uttar Pradesh East, Madhya Pradesh, Bihar, and Delhi) and the remaining states. These four well-represented states all showed less spatial homophily on the busier days. Overall there is moderate negative correlation (Pearson's  $\rho=-0.27$) between $Q_{r}^{D}$ and average logarithmic daily representation at the Kumbh.

\section*{Discussion}

We used CDRs to estimate daily and cumulative attendance at the 2013 Kumbh Mela which, according to our analyses, represents the largest gathering of people in recorded history. While participants from all states demonstrated social and spatial homophily, these phenomena were stronger for the states with low representation at the event and were further amplified on especially crowded days. 

Given that a person may not use their phone immediately upon arriving or before leaving the Kumbh, it is likely that the duration of stay as estimated by their phone usage is truncated. To account for this censoring, a model for daily phone usage is required that can estimate the amount of censoring. We chose the simple model that assumed that each person had some independent probability of using their phone on each day. While this model is intuitive and provides suitable estimates for the amount of censoring, it may be the case that phone usage is captured better by a more complicated and involved model.

Though we consider the proportion of connected triples that are closed in the Kumbh social network as a way of measuring the homophilous tendencies of attendees from each state, we draw a distinction between this measure and what is more commonly known as triadic closure. In the social network context, triadic closure is the mechanism by which connections are formed through a mutual acquaintance. However, since we do not observe when the original network ties are formed, we cannot comment on triadic closure \cite{simmel1950sociology} as a causal mechanism for tie formation. Our observations avoid a causal connotation and focus instead on observed associative measures.

Our finding on spatial homophily is compatible with the phenomenon of ``associative homophily,'' which states that at a social gathering a person is more likely to join or continue engagement with a group as long as that group contains at least one other person who is similar to her \cite{ingram2007people}. Because every group is likely to have at least one person from the majority, associative homophily plays a relatively weak role for someone in the majority as she will be comfortable in almost every group. On the other hand, a person in the minority may have to actively find a group that contains another person similar to him, inflating the minority group's apparent homophily. This framework offers one possible explanation for the tighter cohesion of the states at the Kumbh with low representation.

In conclusion, whether at the individual, group, or state level, it appears that no one likes to be outnumbered. We all seek safety in numbers.

\section*{Supporting Information}


\subsection*{S1 Text}
\label{S1_text}
{\bf Supplementary Text.}  Extended discussion of how some measures of homophily can be susceptible to confounding with the size of the subgroup. The names and corresponding market shares of the 23 mainland states of India is listed. Some intuition for how censoring takes effect is also included.

\vspace{1pc}\noindent\textbf{Figure S1. Stochastic block model edge probabilities by state.} The $p_{kk}$ represent the probability that any random two nodes in state $k$ share an edge, assuming this probability is the same for all pairs of nodes in state $k$. The strong association between this probability and state representation is heavily biased under model mispecification as is more likely the case here, exagerating the result. The baseline probability is calculated assuming no block structure, i.e. all nodes have the same probability of being connected to one another regardless of state membership.
\vspace{1pc}

\vspace{1pc}\noindent\textbf{Figure S2. Simple illustration of the bias produced by the stochastic block model under model misspecification.} Social groups are displayed in blue, and are assumed to all be of equal size. The probability that two people in the same social group share an edge is $0.20$. The probability that two people in different social groups share an edge is $0.04$. 
States A and B are constructed to have identical homophily, i.e. the probability of an edge between two people in the same social group is the same for both states. The average edge probability displayed takes the average over all possible pairs of nodes in the state.
\vspace{1pc}

\vspace{1pc}\noindent\textbf{Figure S3. Schematic for estimation of the probability of phone usage on any given day.} Each square represents a different day, and it is assumed that a person arrives at and departs from the Kumbh only once. The estimated proportion of days a phone is used is calculated as the total number days a phone is used summed across all customers, divided by the length of stay summed across all customers.
\vspace{1pc}

\vspace{1pc}\noindent\textbf{Table S1. State Acronyms and operator market share.} The acronyms for the twenty-three telecommunications states in India used by the operator are listed. In addition, the market share of the operator as measured by the percentage of the total number of people in the state with some form of subscription to a phone plan, taken from the month of January 2013.
\vspace{1pc}

\subsection*{S2 Information}
\label{S2_information}
{\bf Supplementary Information.}  Network data taken over the full duration of the Kumbh Mela.  Daily handset count data, stratified by state.

\section*{Acknowledgments}
IB and JPO are supported by Harvard T.H. Chan School of Public Health Career Incubator Award to JPO. TK is supported by the HBS Division of Research. The authors declare no conflict of interests. The authors would like to thank Gautam Ahuja, Clare Evans, Gokul Madhavan, Daniel Malter, and Peter Sloot for contributing their helpful comments, suggestions, critiques and discussion, and would like to thank the operator for providing access to their data. A special thanks to the operator for both providing us access to their data as well as accomodating us on their campus grounds as we worked on the analysis. In particular, we wish to express our thanks to employees Rohit Dev and Vikas Singhal for their assistance.

%
%
%

\clearpage

\begin{figure}
\begin{center}
\includegraphics[scale=.8]{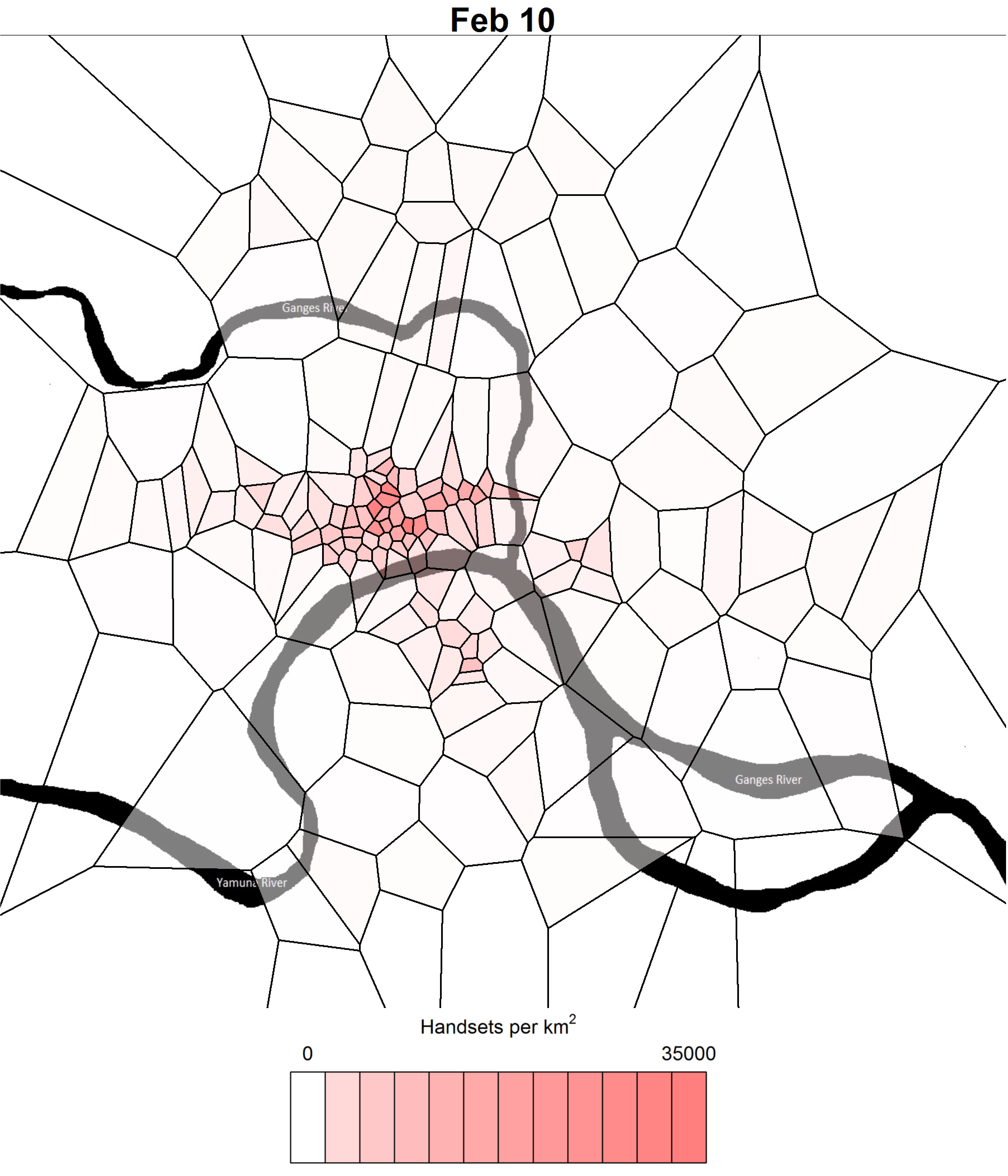}
\caption{}
\end{center}
\end{figure}

\clearpage

\begin{figure}
\begin{center}
\includegraphics[scale=.9]{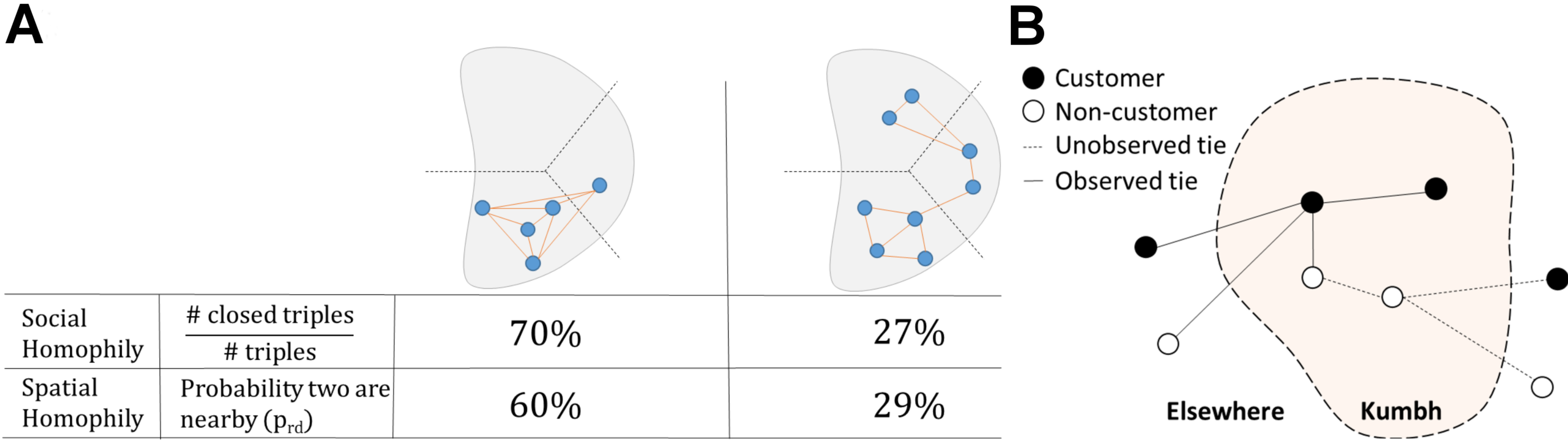}
\caption{}
\end{center}
\end{figure}

\clearpage

\begin{figure}
\begin{center}
\includegraphics[scale=.9]{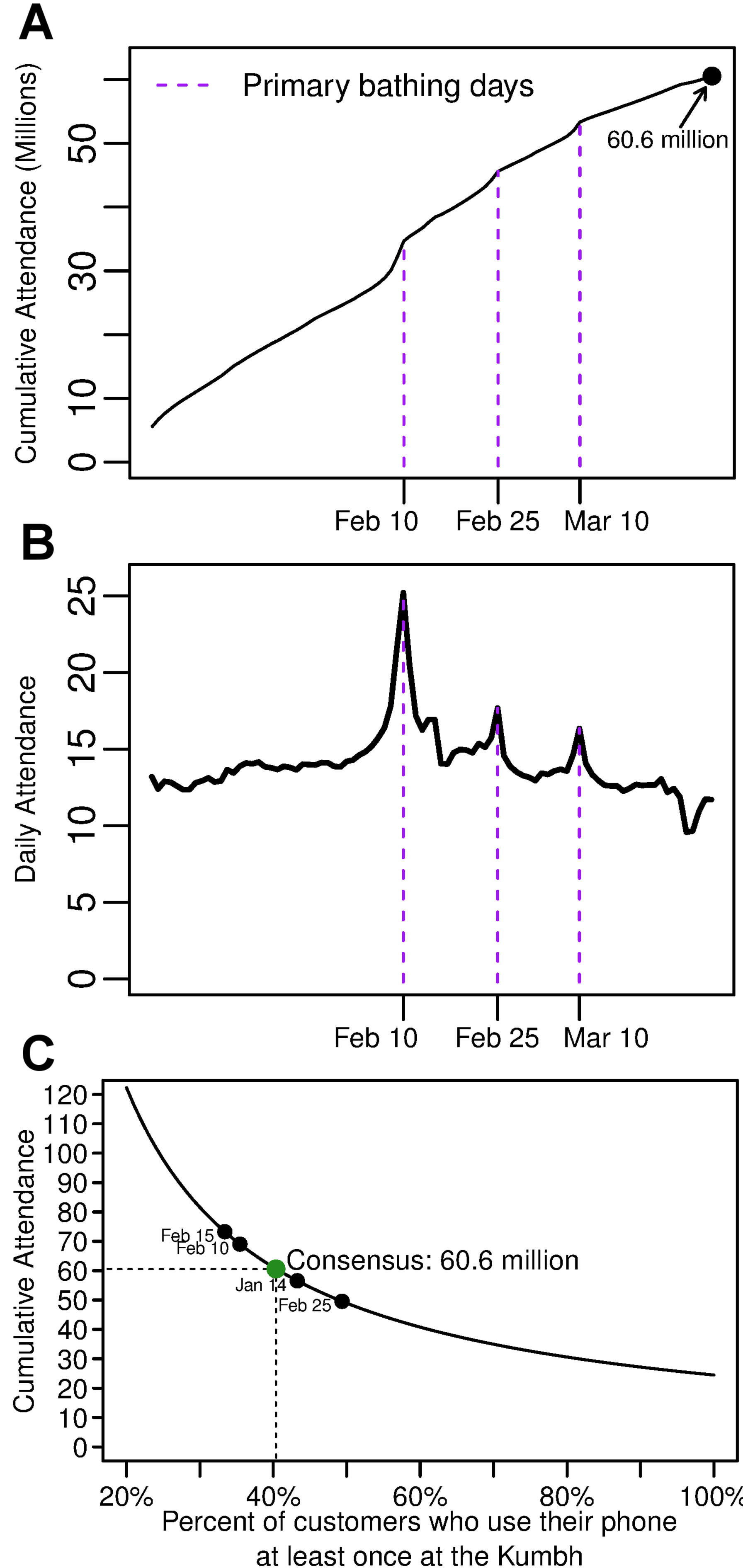}
\caption{}
\end{center}
\end{figure}

\clearpage

\begin{figure}
\begin{center}
\includegraphics[scale=.9]{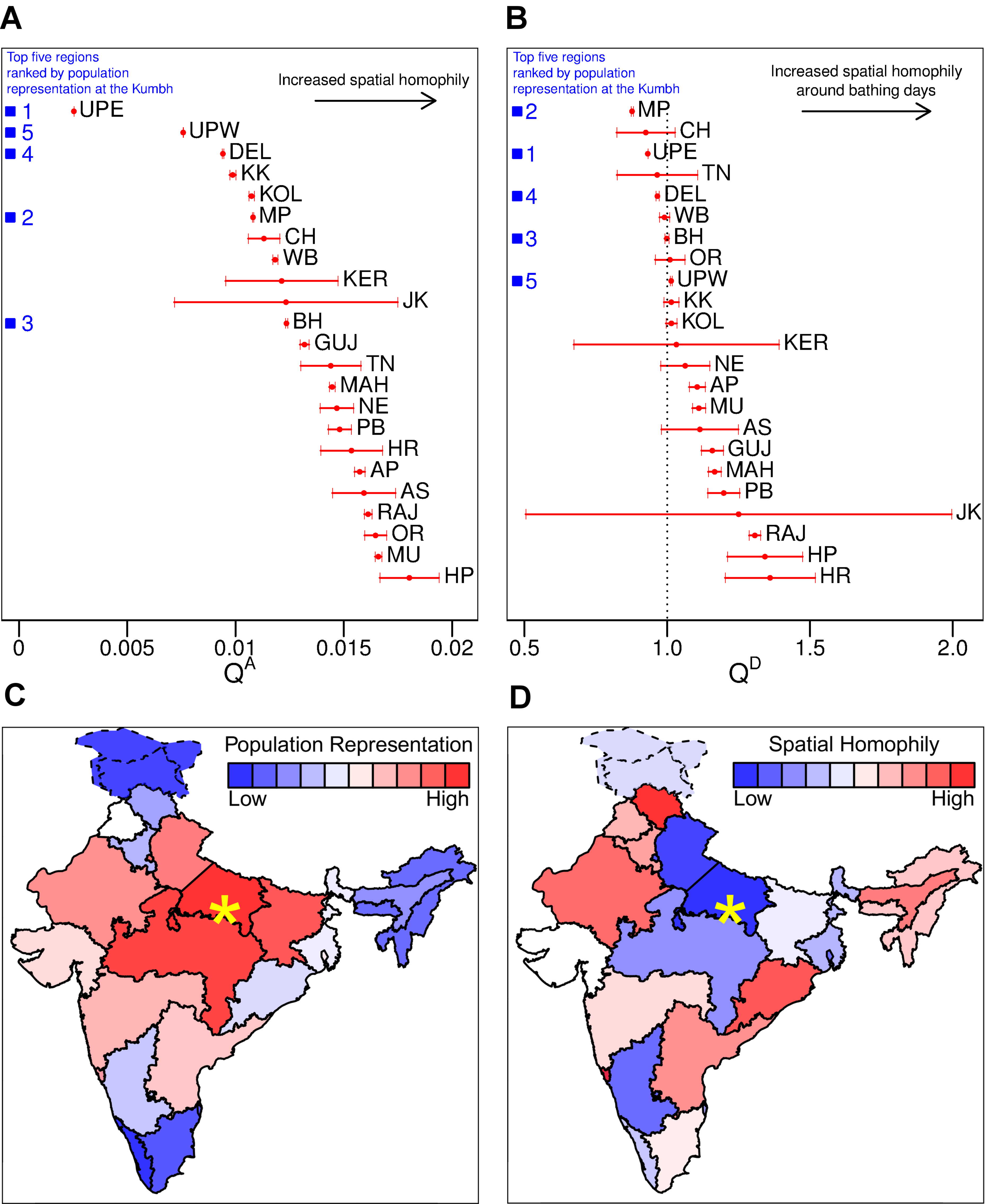}
\caption{}
\end{center}
\end{figure}

\end{document}